\documentclass{JHEP3}

\usepackage{amsfonts}
\usepackage{amsthm}
\usepackage{amssymb}
\usepackage{amscd}
\usepackage{graphicx}

\newcommand{\be}{\begin{equation}}
\newcommand{\ee}{\end{equation}}
\newcommand{\bea}{\begin{eqnarray}}
\newcommand{\eea}{\end{eqnarray}}
\newcommand{\A}{{\mathbf A}}
\newcommand{\X}{{\mathbf X}}
\newcommand{\pro}{\partial}

\newcommand{\V}{{\mathbf V}}

\newcommand{\B}{{\mathbf B}}

\newcommand{\nn}{\nonumber}

\newcommand{\hn}{{\mathbf{\hat n}}}

\newcommand{\hbn}{{\bar n}}
\newcommand{\fbar}{{\bar f}}

\newcommand{\half}{\frac{1}{2}}

\title{Stability of the Magnetic Monopole Condensate in three- and four-colour QCD}
\author{Michael L. Walker\\
Department of Physics, Faculty of Science, Chiba University, Chiba, 263-8522 Japan\\ E-mail: \email{m.walker@aip.org.au}}

\abstract{It is argued
that the ground state of three- and four-colour QCD contains a monopole condensate,
necessary for the dual Meissner effect to be the mechanism of confinement, and
support its stability on the grounds that it gives the off-diagonal gluons an effective mass sufficient to remove the unstable ground state mode.}

\preprint{Chiba Univ.\ Preprint CHIBA-EP-159}

\begin{document}

\section{Introduction}
Proof of colour confinement is one of the most important, long-running 
problems in quantum field theory today. Thanks to the efforts of many authors, 
such as \cite{EI82,S88,SST95,Kondo98a,F02,CP02}, we may now be close to solving this puzzle. A particularly promising
mechanism is the dual Meissner effect, in which a condensate of chromomagnetic 
monopoles excludes the chromoelectric field analogously to the Cooper pairs in
a superconductor excluding the magnetic field. This proposal,
dating back to the middle 1970s \cite{N74,M76,P77,tH81,Cho80a}, 
requires QCD to have a magnetic monopole 
condensate. One obvious difficulty
was ensuring that the magnetic condensate was due to monopoles, but the most
discouraging was the result of Nielsen and Olesen \cite{NO78} in two-colour QCD
that a magnetic condensate renders the zero-point gluon fluctuations unstable. 
Although this instability was disputed \cite{H72,S82,DR83,F83b}, its existence 
remained conventional wisdom until relatively recently.
Cho \textit{et. al.}, using subtle causality
considerations, have argued that Nielsen and Olesen's analysis was too 
n$\ddot{\mbox{a}}$ive and 
found instead that the imaginary part of the effective action was zero for magnetic
backgrounds but non-zero for electric backgrounds \cite{CP02}. Together with the 
current author, they have supported
their result with independent calculations \cite{CmeP04,Cme04}, and recently
extended it to three or more colours \cite{CKP05}.
A different approach, taken by Kondo \cite{K04} in two-colour QCD, demonstrates the 
generation
of an effective gluon mass large enough to remove the tachyon mode. We shall see that
this argument has parallels with that of Flory \cite{F83b} and Kay \textit{et.al.} \cite{KKP05}.

I shall repeat Kondo's approach in three- and four-colour QCD. Section \ref{sec:CFN}
presents the Cho-Faddeev-Niemi (CFN) decomposition for general $SU(N)$
gauge groups. Section \ref{sec:QCDvac} determines
the magnetic and monopole condensates, drawing heavily on the maximal abelian gauge analysis of
Flyvbjerg \cite{F80}. This is followed by a discussion of the monopole generating 
subgroups $U(1)^{N-1}$ and $U(N-1)$ and the different roles they play confining
gluons or quarks in section \ref{sec:subgroup}. I study the
gluon's effective mass matrix and determine the effective mass (squared) 
in terms of the magnitude of the monopole field in
section \ref{sec:gluonmass}. The apparent instability is briefly discussed in 
section \ref{sec:issue}. I establish inequalities between the magnetic
and monopole condensates in section \ref{sec:twokinds}.
Section \ref{sec:SU4stability} adapts this approach to four-colour QCD. 

\section{Specifying Abelian Directions} \label{sec:CFN}
The CFN decomposition was first presented by Cho \cite{Cho80a}, and later by Faddeev
and Niemi \cite{FN99}, as a gauge-invariant means of specifying
the Abelian dynamics of two-colour QCD. These authors \cite{C80,FN99b}
also applied it to three-colour QCD. In this section we adapt it to general
$SU(N)$, although we are not the first to do so \cite{FN99c,LZZ00}, and establish our 
notation. 

The Lie group $SU(N)$ for $N$-colour QCD 
has $N^2-1$ generators $\lambda^{(i)}$, of which $N-1$
are Abelian generators $\Lambda^{(i)}$. For simplicity, we specify the 
gauge transformed Abelian directions with $\hn_i = U^\dagger \Lambda^{(i)} U$. 
Fluctuations in the $\hn_i$ directions are described by $c^{(i)}_\mu$. 
The gauge field of the covariant derivative which leaves
the $\hn_i$ invariant is given by
\bea
g\mathbf{V}_\mu \times \hn_i = -\partial_\mu \hn_i
\eea
In general this is 
\bea
\mathbf{V}_\mu = c^{(i)}_\mu \hn_i + \mathbf{B}_\mu ,\;
\mathbf{B}_\mu = g^{-1} \partial_\mu \hn_i \times \hn_i,
\eea
where summation is implied over $i$.

We define the covariant derivative
\be
\mathbf{\hat{D}}_\mu = \partial_\mu + g\mathbf{V}_\mu \times
\ee

The monopole field strength
\be
\vec{H}_{\mu \nu} = \partial_\mu \mathbf{B}_\nu - \partial_\nu \mathbf{B}_\mu
+ g\mathbf{B}_\mu \times \mathbf{B}_\nu,
\ee
has only $\hn_i$ components, \textit{ie}. 
\be
H^{(i)}_{\mu\nu}\hn_i = \vec{H}_{\mu\nu},
\ee
where $H^{(i)}_{\mu\nu}$ has the eigenvalue $H^{(i)}$. Since we are only
concerned with magnetic backgrounds, $H^{(i)}$ is considered the magnitude
of a background magnetic field $\mathbf{H}^{(i)}$.

$\X_\mu$ is defined to be the dynamical degrees of freedom (DOF) perpendicular to $\hn_i$, 
so if $\A_\mu$ is the gluon field then
\bea
\A_\mu &=& \V_\mu + \X_\mu = c^{(i)}_\mu \hn_i + \mathbf{B}_\mu + \X_\mu,
\eea
where
\bea \label{eq:Xdefn}
\X_\mu &\bot& \hn_i ,\;
\X_\mu = g^{-1}\hn_i \times \mathbf{D}_\mu \hn_i,\;
\mathbf{D}_\mu = \partial_\mu + g\A_\mu \times. 
\eea

Substituting the CFN decomposition into the QCD field strength tensor gives
\bea
\vec{F}^2 &=&
(\pro_\mu c^{(i)}_\nu - \pro_\nu c^{(i)}_\mu)^2 
+(\pro_\mu \B_\nu - \pro_\nu \B_\mu + g \B_\mu \times \B_\nu)^2 \nn \\
&&+ 2(\pro_\mu c^{(i)}_\nu - \pro_\nu c^{(i)}_\mu)\hn_i 
\cdot (\pro_\mu \B_\nu - \pro_\nu \B_\mu + g \B_\mu \times \B_\nu) 
+ (\mathbf{\hat{D}}_\mu \X_\nu - \mathbf{\hat{D}}_\nu \X_\mu)^2 \nn \\
&& +2g((\pro_\mu c^{(i)}_\nu - \pro_\nu c^{(i)}_\mu)\hn_i 
+ \pro_\mu \B_\nu - \pro_\nu \B_\mu + g \B_\mu \times \B_\nu) 
\cdot (\X_\mu \times \X_\nu) \nn \\
&& +g^2 (\X_\mu \times \X_\nu)^2 
+ 2g (\mathbf{\hat{D}}_\mu \X_\nu - \mathbf{\hat{D}}_\nu \X_\mu) \cdot (\X_\mu \times \X_\nu).
\eea
This expression holds for all $N$-colour QCD except $N=2$ where the last term is 
absent.

A n$\ddot{\mbox{a}}$ive substitution of the CFN decomposition appears to leave the gluon field with additional DOF, and this has been a source of considerable confusion
and controversy. Detailed analyses can be found in
\cite{BCK02,KMS05,K06} demonstrating that 
the $\hn_i$ are not fundamental, but a compound of dynamic fields.
Hence $\hn_i, \B_\mu$ are dynamic but do not constitute extra DOFs.

However the CFN decomposition does introduce additional gauge DOFs, which a 
proper application must fix. \cite{CP02,BCK02} discussed the problem effectively in 
terms of the passive and active gauge symmetries, but I shall follow the
notation of \cite{KMS05}. Their analysis was restricted to two-colour QCD, but its application to $N$-colours is so straightforward
as to be little more than repetition. It is sufficient for our purposes to say that 
the CFN decomposition of QCD can be properly quantised in a consistent manner that
leaves it equivalent to conventional QCD.

\section{The $SU(3)$ CFN QCD Vacuum} \label{sec:QCDvac}
To discuss the vacuum state we employ the formalism of Lie algebra roots to the isovectors
$\B_\mu, \X_\mu$, reducing them to 
\bea \label{eq:roots}
\B_\mu &=& \B_\mu^{(1,0)} + \B_\mu^{(\half, \frac{\sqrt{3}}{2})}
+ \B_\mu^{(\half, -\frac{\sqrt{3}}{2})} \;\;
= \B_\mu^{(1,0)} + \B_\mu^{(-\half, -\frac{\sqrt{3}}{2})}
+ \B_\mu^{(-\half, \frac{\sqrt{3}}{2})} ,\nn \\
\X_\mu &=& \X_\mu^{(1,0)} + \X_\mu^{(\half, \frac{\sqrt{3}}{2})}
+ \X_\mu^{(\half, -\frac{\sqrt{3}}{2})} \;\;
= \X_\mu^{(1,0)} + \X_\mu^{(-\half, -\frac{\sqrt{3}}{2})}
+ \X_\mu^{(-\half, \frac{\sqrt{3}}{2})}. \nn \\
\eea
$\B^{(\alpha)}_\mu$ is defined so that 
\be
g\B^{(\alpha)}_\mu \times \B^{(\alpha)}_\nu = - \vec{H}_{\mu \nu}^{(\alpha)} ,
\ee
while $\mathbf{X}_\mu^{(\alpha)}$ is the component of $\X_\mu$ which feels the monopole
field strength tensor $\vec{H}_{\mu \nu}^{(\alpha)}$, where
\be
\vec{H}_{\mu \nu}^{(\alpha)} = \alpha_j H_{\mu \nu}^{(j)}.
\ee
We also define the background magnetic field
\be
\mathbf{H}^{(\alpha)} = \alpha_j \mathbf{H}^{(j)},
\ee
whose magnitude $H^{(\alpha)}$ is $\vec{H}_{\mu \nu}^{(\alpha)}$'s non-zero eigenvalue. It follows
that
\bea
H^{(1,0)} &=& H^{(1)}, \;\;
H^{(\half, \pm \frac{\sqrt{3}}{2})^2}
= \frac{1}{4} H^{(1)^2} + \frac{3}{4} H^{(2)^2}
\pm \frac{\sqrt{3}}{2} \mathbf{H}^{(1)} \cdot \mathbf{H}^{(2)} .
\eea

This result is formally the same as Flyvberg's \cite{F80}, with the subtle difference
that our $H^{(\alpha)}$ refers to the the field strength generated by the Cho
connection while Flyvberg's is simply the field strength along the Abelian directions
in the maximal Abelian gauge. Nonetheless, it is clear that we can repeat
the renormalization analysis and get the same formal result. This gives the corresponding results for the lowest energy state,
\bea \label{eq:QCDvacuum}
H^{(1)} =  H^{(2)}, \;\;
\mathbf{H}^{(1)} &\bot & \mathbf{H}^{(2)},
\eea
as found independently for the CFN formalism 
using a different approach by Cho, Kim and Pak \cite{CKP05}. 

\section{$U(1)^{N-1}$ Monopoles vs $U(N-1)$ Monopoles} \label{sec:subgroup}
Since the ultimate motivation of this work is confinement, it is appropriate
to discuss an important issue first brought to light by
Kondo and Taira \cite{KT00a,KT00b} in their construction of a non-Abelian version
of Stokes' theorem. They found that the monopole contribution to the Wilson
loop depends on which representation of the gauge group the colour charge belongs
to. In the $SU(3)$ gauge group for example, discussion of the fundamental representation  concerns only the monopoles corresponding to the reduction from $SU(3)$
down to $U(2)$ symmetry \cite{KT00a}, specified by the homotopy group
\be
\pi_2[SU(3)/U(2)] = \pi_1[U(2)] = \pi_1[SU(2)\otimes U_2(1)]
= \pi_1[U_2(1)] = \mathbb{Z}_2,
\ee
(Subscripts $_i$ in this section denote the relevant Abelian generator 
$\Lambda^{(i)}$.)
while for colour charges in the adjoint representation we need to consider
the corresponding $U(1) \otimes U(1)$ fundamental group
\be
\pi_2[SU(3)/(U_1(1)\otimes U_2(1))] = \pi_1[U_1(1)\otimes U_2(1)]
= \mathbb{Z}_1 \oplus \mathbb{Z}_2.
\ee
The Abelian generator $\Lambda^{(1)}$ of the subgroup $U(2)$ is
contained in the simply connected subgroup $SU(2)$, 
leaving only the fundamental group generated by $\Lambda^{(2)}$.
Hence the monopole charges corresponding to the $U(2)$ subgroup are a subset of 
those corresponding to $U_1(1) \otimes U_2(1)$. Specifically, it is the subset
for which the charge corresponding to $U_1(1)$ is zero. 

To construct the monopole field due to the $U(2)$ subgroup, observe that
since the $U(2)$ monopole field is purely $\Lambda^{(2)}$-like it will be the 
covariant connection of the unit vector $\hn_2$. It is easy to show that
\be \label{eq:U2monopole}
\mathbf{L}_\mu = g^{-1} \frac{4}{3}\partial_\mu \hn_2 \times \hn_2,
\ee
has the required property
\be
g\mathbf{L}_\mu \times \hn_2 = - \partial_\mu \hn_2.
\ee
Of course, the $U_1(1) \otimes U_2(1)$ monopole field
$\B_\mu$ also has this property, so
\bea
g(\B_\mu - \mathbf{L}_\mu) \times \hn_2 = 0.
\eea
Since
\bea
g\B_\mu^{(1,0)} \times \hn_2 = 0,
\eea
it follows that
\bea
\mathbf{L}_\mu &=& \B_\mu^{(\half,\frac{\sqrt{3}}{2})} + \B_\mu^{(\half,-\frac{\sqrt{3}}{2})} 
= \B_\mu^{(-\half,-\frac{\sqrt{3}}{2})} + \B_\mu^{(-\half,\frac{\sqrt{3}}{2})}.
\eea

For general $SU(N)$, the Wilson loop for gluons, which belong to the adjoint representation, depends on the full set of monopoles corresponding to the homotopy group
\bea
\pi_2[SU(N)/(U_1(1)^{N-1})] 
= \pi_1[U_1(1)^{N-1}] 
= \mathbb{Z}_1 \oplus \mathbb{Z}_2 \oplus\ldots \oplus\mathbb{Z}_{N-1}.
\eea
This is in contrast to that of quarks in the fundamental representation, which 
receives monopole contributions only from those corresponding to 
\be
\pi_2[SU(N)/U(N-1)] = \pi_1[U(N-1)]
= \pi_1[U_{N-1}(1)] = \mathbb{Z}_{N-1}.
\ee
The covariant connection of the unit vector $\hn_{N-1}$ has the general form
\be 
\mathbf{L}_\mu = g^{-1} K(N)\partial_\mu \hn_{N-1} \times \hn_{N-1}.
\ee
It is now trivial to generalise the $SU(3)$ statement
\bea
(\B_\mu - \mathbf{L}_\mu) \times \hn_{N-1} = 0.
\eea

This section demonstrates that a whole new analysis is unnecessary if quark, rather
than gluon confinement is of interest. Of course, stability of the ground-state fluctuations
of the quark field was never an issue. In the expected absence of internal 
anisotropy, all other results concerning condensates of $\B_\mu$ should also hold for
$\mathbf{L}_\mu$. 

\section{Mass of Off-diagonal $SU(3)$ Gluons} \label{sec:gluonmass}
Following Kondo \cite{K04}, we observe at the classical level that
the monopole condensate gives the off-diagonal gluons an effective mass via
\bea 
\half (\mathbf{\hat{D}}_\mu \X_\nu - \mathbf{\hat{D}}_\nu \X_\mu)^2 
\stackrel{IBP}{\longrightarrow}& (\X_\mu \mathbf{\hat{D}}_\nu)\cdot (\mathbf{\hat{D}}_\mu \X_\nu)
- (\X_\mu \mathbf{\hat{D}}_\nu)\cdot (\mathbf{\hat{D}}_\nu \X_\mu).
\eea
The latter term gives
\be \label{eq:massterm}
\lefteqn{g^2 B_\rho^D X^E_\mu B_\rho^B X_\mu^C f_{ABC} f_{ADE}},
\ee
which provides the effective gluon mass matrix
\bea \label{eq:gluonmass}
M^2_{EC} = g^2 \lefteqn{B_\rho^D B_\rho^B f_{ABC} f_{ADE}}.
\eea
Since the effective
mass term arises from the quartic gluon terms, this is consistent with an early 
calculation by Flory \cite{F83b} and Kay \textit{et.al.} \cite{KKP05} showing that the
instability is removed when the quartic terms relevant to the unstable modes are 
included. Dudal and coworkers are following an entirely different approach 
\cite{DGLSSSV04,DVBG03} in which the 
gluon mass comes from a ghost-gluon condensate with dimensions of mass squared.

So far this section has followed the corresponding section 2.2 in \cite{K04}. 
Because the algebra of $SU(2)$ is simpler than
that of $SU(3)$, the author was able to simply diagonalize the mass
matrix and obtain the mass squared eigenvalues $\B\cdot \B$ (multiplicity two) and
zero. The zero eigenvalue corresponds to the Abelian direction.

Diagonalizing (\ref{eq:gluonmass}) however, is too difficult even for
mathematica but there is another way.
The sum of the mass eigenvalues is the trace of the mass matrix, 
$3g^2 \B\cdot \B$.
Since there are two Abelian directions from which the valence gluon is excluded
by definition (see (\ref{eq:Xdefn})),
it follows that zero is an eigenvalue of multiplicity two and 
the average effective mass squared is
\be \label{eq:avmass}
M_X^2 = \frac{3}{8-2} g^2 \B\cdot \B = \half g^2 \B\cdot \B.
\ee
Since all physical masses are equal by the isotropy of the condensate and the
gauge invariance of the mass term (\ref{eq:massterm}), (\ref{eq:avmass}) is the
effective mass of all valence gluons.
A conventional diagonalization of $M^2_{EC}$ in this treatment would, of course
have been preferable, but this approach does
give the same result as diagonalization in $SU(2)$ QCD.

\section{Is the Monopole Condensate Stable in $SU(3)$ QCD?} \label{sec:issue}
It has been shown \cite{F80,S77} that
\bea
\Vert \mathbf{H}^{(\alpha)} \Vert \ne 0 ,
\eea
but a calculation of the
$\X^{(\alpha)}$ ground-state energy using zeta-function renormalization, as first demonstrated in two-colour QCD~\cite{NO78}
by Nielsen and Olesen, has an imaginary contribution \cite{meK06} from
\bea
\sqrt{\mathbf{k}^2 - g\Vert\mathbf{H}^{(\alpha)}\Vert}.
\eea
However there is still hope, because we saw in section \ref{sec:gluonmass} 
that the gluons gain an effective mass, changing this to
\bea
\sqrt{\mathbf{k}^2 + M_X^2 - g\Vert\mathbf{H}^{(\alpha)}\Vert}.
\eea
It now remains to demonstate that the spin contribution is smaller in magnitude than the effective gluon mass squared.

\section{Monopole vs the Magnetic Condensate} \label{sec:twokinds}
Since $\langle \Vert H^{(\alpha)} \Vert \rangle$ does not vary with $\alpha$,
proving sufficient $M_X^2$ to prevent tachyons 
for $\X^{(1,0)}_\mu$ is sufficient to prove it for $\X_\mu$.
Noting
\bea
(\B^{(\alpha)}_\mu \times \B^{(\alpha)}_\nu)^2 
= (\B^{(\alpha)}_\mu \cdot \B^{(\alpha)}_\mu)^2 
-(\B^{(\alpha)}_\mu \cdot \B^{(\alpha)}_\nu)^2.
\eea
gives
\bea \label{eq:H1inequal}
\lefteqn{\Vert g\mathbf{H}^{(1,0)} \Vert} \nn \\
&=& g^2 \Vert \hn_1 \cdot \B^{(1,0)}_\mu \times \B^{(1,0)}_\nu 
+ \hn_1 \cdot \B^{(\half,\frac{\sqrt{3}}{2})}_\mu \times \B^{(\half,\frac{\sqrt{3}}{2})}_\nu 
+ \hn_1 \cdot \B^{(\half,-\frac{\sqrt{3}}{2})}_\mu \times \B^{(\half,-\frac{\sqrt{3}}{2})}_\nu \Vert
\nn \\
&\le& \frac{2g^2}{3} \B \cdot \B,
\eea
which is not strong enough. We remedy this by showing that
\be \label{eq:sqrt2}
\sqrt{2} \Vert \B^{(\alpha)}_\mu \times \B^{(\alpha)}_\nu \Vert 
\le \B^{(\alpha)}_\mu \cdot \B^{(\alpha)}_\mu.
\ee
Begin by constructing a convenient coordinate system.
Let $\{\mathbf{\bar{n}}_i \}_{i=1}^{N^2-1}$ be unit vectors spanning $SU(N)$ .
$\mathbf{\bar{n}}_i \cdot \mathbf{\bar{n}}_j \times \mathbf{\bar{n}}_k$ 
is gauge invariant under
\begin{displaymath}
\delta \mathbf{\bar{n}}_i = \mathbf{\bar{n}}_i \times \mathbf{\alpha}, 
\end{displaymath}
so
\be
\mathbf{\bar{n}}_i \cdot \mathbf{\bar{n}}_j \times \mathbf{\bar{n}}_k \equiv \bar{f}_{ijk} = f_{ijk}.
\ee
We can construct a convenient coordinate system by starting with 
\be
\{\mathbf{\bar{n}}_i \} = \{\mathbf{\hat{e}}_i \},
\ee
and gauge transforming $\{\mathbf{\bar{n}}_i \}$ so that 
\be
\mathbf{\bar{n}}_3 , \mathbf{\bar{n}}_8 = \hn_1,\hn_2,
\ee
respectively. It follows that
\bea
\bar{f}_{ijk} = f_{ijk}.
\eea
Restricting the analysis to $SU(3)$, $\B_\mu^{\pm(1,0)}, \X_\mu^{\pm(1,0)}$ lie in the
$\{\mathbf{\bar{n}}_1,\mathbf{\bar{n}}_2 \}$ plane,  
$\B_\mu^{\pm(\half,\frac{\sqrt{3}}{2})}, \X_\mu^{\pm(\half,\frac{\sqrt{3}}{2})}$
lie in the $\{\mathbf{\bar{n}}_4,\mathbf{\bar{n}}_5 \}$ plane, and $\B_\mu^{\pm(\half,-\frac{\sqrt{3}}{2})}, \X_\mu^{\pm(\half,-\frac{\sqrt{3}}{2})}$
lie in the $\{\mathbf{\bar{n}}_6,\mathbf{\bar{n}}_7 \}$ plane.

The following is based on a method developed by Kondo \cite{Kbook06} for
two-colour QCD. Define
\be
T^{ab} = \hbn^a \cdot \partial \hn_1 \; \hbn^b \cdot \partial \hn_1
\ee
where $a,b$ are restricted to $1,2$. 
$T^{ab}$ is a two by two matrix, having two real
eigenvalues, $\lambda_1$ and $\lambda_2$ say. We find the inequality
\be \label{eq:sqrt2inequal}
\half \left(\sum_{a=1}^2 \lambda_a \right)^2
= \lambda_1^2 + \lambda_2^2 - \half(\lambda_1 - \lambda_2)^2
\le \sum_{a=1}^2 \lambda_a^2
\rightarrow \half (\mbox{Tr}\, T)^2 \le (\mbox{Tr} \, T^2).
\ee
Proof of (\ref{eq:sqrt2}) for $H^{(1,0)}_{\mu \nu}$ is straightforward.
Take 
\bea
[\partial_\mu \hn_1]^a = \hbn^a \, \hbn^a \cdot \partial_\mu \hn_1 \;
\mbox{(no summation),}
\eea
where $a$ is restricted to $1,2$. We get
\bea 
g^2\vec{H}^{(1,0)}_{\mu \nu} \cdot \vec{H}^{(1,0)}_{\mu \nu} &=&
g^4(\B_\mu^{(1,0)} \times \B_\nu^{(1,0)})^2 
= \fbar^{3ab} [\partial_\mu \hn_1 ]^a [\partial_\nu \hn_1 ]^b 
\fbar^{3cd} [\partial_\mu \hn_1 ]^c [\partial_\nu \hn_1 ]^d \nn \\
&=& (\mbox{Tr} \, T)^2 - (\mbox{Tr} \, T^2).
\eea
Substituting in (\ref{eq:sqrt2inequal}) we find 
\bea
\vec{H}^{(1,0)}_{\mu \nu} \cdot \vec{H}^{(1,0)}_{\mu \nu} \le 
\half g^2(\B_\mu^{(1,0)} \cdot \B_\mu^{(1,0)})^2,
\eea
which leads to equation (\ref{eq:sqrt2}).
The construction for $\vec{H}^{\pm(\half,\frac{\sqrt{3}}{2})}$ is only
slightly more complicated. Redefine
\be
T^{(a-3)(b-3)} = \hbn^a \cdot \partial \hn_2 \; \hbn^b \cdot \partial \hn_2
\ee
where $a,b$ are restricted to $4,5$. Now take 
\bea
[\partial_\mu \hn_2]^a = \hbn^a \, \hbn^a \cdot \partial_\mu \hn_2 ,
\eea
where $a$ is still restricted to $4,5$. Recalling the discussion of equation 
(\ref{eq:U2monopole}) and repeating the above argument leads to
\bea 
g^2\lefteqn{\vec{H}^{\pm\Big(\half,\frac{\sqrt{3}}{2}\Big)}_{\mu \nu} \cdot \vec{H}^{\pm\Big(\half,\frac{\sqrt{3}}{2}\Big)}_{\mu \nu} =
g^4\Big(\B_\mu^{\pm\Big(\half,\frac{\sqrt{3}}{2}\Big)} \times \B_\nu^{\pm\Big(\half,\frac{\sqrt{3}}{2}\Big)}\Big)^2} \nn \\
&=& \frac{16}{9}\Big(\frac{1}{4} \fbar^{3ab} \fbar^{3cd} 
+ \frac{3}{4}  \fbar^{8ab} \fbar^{8cd} \Big) 
[\partial_\mu \hn_2 ]^a [\partial_\nu \hn_2 ]^b 
[\partial_\mu \hn_2 ]^c [\partial_\nu \hn_2 ]^d  \nn \\
&=& \frac{16}{9}((\mbox{Tr} \, T)^2 - (\mbox{Tr} \, T^2)),
\eea
which again yields equation (\ref{eq:sqrt2}).
The argument for $\vec{H}^{\pm\Big(\half,\frac{\sqrt{3}}{2}\Big)}$ is identical.
The adaptation of this technique to higher $N$ is straightforward.

The above introduces a factor of $\sqrt{2}$ to the 
inequality (\ref{eq:H1inequal}), which 
becomes
\bea
\lefteqn{\Vert g\mathbf{H}^{(1,0)} \Vert} \nn \\ 
&=& g^2 \Big\Vert \hn_1 \cdot \B^{(1,0)}_\mu \times \B^{(1,0)}_\nu 
+ \hn_1 \cdot \B^{(\half,\frac{\sqrt{3}}{2})}_\mu \times \B^{(\half,\frac{\sqrt{3}}{2})}_\nu 
+ \hn_1 \cdot \B^{(\half,-\frac{\sqrt{3}}{2})}_\mu \times \B^{(\half,-\frac{\sqrt{3}}{2})}_\nu \Big\Vert
\nn \\
&\le& \frac{2g^2}{3\sqrt{2}} \B \cdot \B < \half g^2\B \cdot \B,
\eea
demonstrating that the effective mass is sufficient to stabilize the tachyonic gluon mode.

While unnecessary for $SU(3)$, it is possible to use 
$\langle \Vert H^{(1)} \Vert \rangle = \langle \Vert H^{(2)} \Vert \rangle$ to
find an even stronger upper bound on $\langle \Vert \vec{H}^{(\alpha)} \Vert \rangle$.
\bea
\Vert g\mathbf{H}^{(2)} \Vert \le g^2\frac{\sqrt{3}}{2} 
\Big(\Big\Vert \B^{\Big(\half, \frac{\sqrt{3}}{2}\Big)}_\mu 
\times \B^{\Big(\half, \frac{\sqrt{3}}{2}\Big)}_\nu 
+ \B^{\Big(\half, -\frac{\sqrt{3}}{2}\Big)}_\mu 
\times \B^{\Big(\half, -\frac{\sqrt{3}}{2}\Big)}_\nu \Big\Vert\Big),
\eea
yielding
\bea
\Vert g\mathbf{H}^{(2)} \Vert \le g^2 \frac{\sqrt{3}}{3\sqrt{2}} \B \cdot \B
= \frac{g^2}{\sqrt{6}} \B \cdot \B
< \frac{g^2\sqrt{2}}{3}\B \cdot \B.
\eea
This style of argument will prove necessary in the treatment of $SU(4)$.

\section{Stability of $SU(4)$ QCD} \label{sec:SU4stability}
Repeating the analysis of section \ref{sec:gluonmass} finds an effective gluon mass squared
in $SU(4)$ QCD of
\be
M^2_{SU(4)} = \frac{4g^2\B \cdot \B}{12}  = \frac{g^2}{3} \B \cdot \B.
\ee
Following the last section, we again need only one 
$\langle \Vert g\mathbf{H}^{(i)} \Vert \rangle < M^2_{SU(4)}$. For $\mathbf{H}^{(1)}$ we get
\be
\Vert g\mathbf{H}^{(1)} \Vert \le \frac{g^2}{2\sqrt{2}} \B \cdot \B > M^2_{SU(4)}
\ee
but studying $\mathbf{H}^{(3)}$ yields
\bea
\Vert g\mathbf{H}^{(3)} \Vert &\le& \sqrt{\frac{2}{3}}\Big\Vert 
g^2\B^{\Big(\half, \sqrt{\frac{1}{12}}, \sqrt{\frac{2}{3}}\Big)}_\mu
\times \B^{\Big(\half, \sqrt{\frac{1}{12}}, \sqrt{\frac{2}{3}}\Big)}_\nu 
\Big\Vert \nn \\
&& + \sqrt{\frac{2}{3}} \Big\Vert 
g^2\B^{\Big(\half, -\sqrt{\frac{1}{12}}, -\sqrt{\frac{2}{3}}\Big)}_\mu
\times \B^{\Big(\half, -\sqrt{\frac{1}{12}}, -\sqrt{\frac{2}{3}}\Big)}_\nu \Big\Vert \nn \\
&& + \sqrt{\frac{2}{3}} \Big\Vert g^2\B^{(0, \sqrt{\frac{1}{3}}, -\sqrt{\frac{2}{3}}\Big)}_\mu
\times \B^{\Big(0, \sqrt{\frac{1}{3}}, -\sqrt{\frac{2}{3}}\Big)}_\nu \Big\Vert \nn \\
&\le& \frac{g^2}{2\sqrt{3}} \B \cdot \B < M_{SU(4)}^2,
\eea
protecting the monopole condensate in $SU(4)$ QCD. 
It is pointless to try and generalize
this result to arbitrary $SU(N>4)$ in three dimensional space because satisfying the equations (\ref{eq:QCDvacuum}) requires $N-1$ mutually orthogonal vector fields.

\section{Discussion} \label{sec:discussion}
A case for a stable monopole condensate in the QCD vacuum has been presented.
By adapting the CFN decomposition to the higher gauge group we have ensured that our
analysis describes the monopoles in a consistent, gauge invariant manner.
Applying the CFN decomposition to $SU(N>2)$ is 
straightforward and reasonably intuitive. This was also the experience of Cho, Kim and
Pak \cite{CKP05} who have demonstrated condensate stability in $SU(3)$ QCD 
by calculating the imaginary part of the effective action as discussed earlier.

The CFN formalism, while different from 
and superior to t'Hooft's Abelian gauge, has sufficient
formal similarity for Flyvbjerg's analysis \cite{F80} 
to carry over to it, so we inherit the corresponding
results concerning the QCD ground state in section \ref{sec:QCDvac}. 

When discussing whether the effective 
gluon mass is sufficient to stabilize the ground state,
it is important to remember that the relevant magnetic field strength
magnitudes are found in the gluon 
spin interaction $\Vert \mathbf{H}^{(\alpha)} \Vert$. Diagonalizing the mass matrix 
directly seems impossible, but the invariance of the mass-generating term under
global active gauge transformations ensures that the mass eigenvalues are
equal, allowing their deduction from the trace of the mass matrix. It must be 
remembered that the construction of the gluon mass squared matrix was a classical
one, even though a one-loop calculation is providing the non-zero condensate. Complete 
proof requires the mass matrix calculation to be a quantum one. The approach of Dudal
and coworkers \cite{DGLSSSV04,DVBG03} is interesting in this regard.

It has been shown explicitly for three-colour QCD that the CFN decomposition  corresponding to the maximal
Abelian subgroup contains the monopoles corresponding to the $U(2)$ subgroup automatically. It is not hard to 
derive the corresponding result for the CFN decomposition in four-colour QCD 
from the non-trivial homotopy groups for $SU(4)$ \cite{KKP82}.

Our main result is that applying Kondo's argument \cite{K04} to $SU(3)$ or
$SU(4)$ QCD finds an effective gluon mass sufficient to stabilize the monopole 
condensate. $SU(N>4)$ QCD requires a new analysis for
reasons given at the end of section \ref{sec:SU4stability}. 

\acknowledgments{The 
author wishes to thank K.-I. Kondo for many helpful discussions and for 
proof-reading the manuscript. He also thanks T. Shinohara and T. Murakami for 
many helpful discussions and criticisms. The author is supported by a postdoctoral
fellowship from the Japan Society for the Promotion of Science (JSPS)}


\providecommand{\href}[2]{#2}\begingroup\raggedright\endgroup

\end{document}